\documentstyle[aps,epsf,rotate,preprint]{revtex}

\begin{document}

\draft

\centerline{\bf {Erratum: Small-world networks: Evidence for a crossover
picture}}

\centerline{\bf {[Phys. Rev. Lett. 82, 3180 (1999)]}}
\vspace*{1cm}

\centerline{Marc Barth\'el\'emy and Lu\'{\i}s A. Nunes Amaral}

\vspace*{1cm}
We have performed new calculations using the breadth-first search
algorithm [1,2]. We are now able to study systems with sizes
up to $n=5500$. As shown in Fig.~1, we now find $\tau\approx 1$, in
agreement with the simple argument given in our Letter but different
from the originally reported numerical result ($\tau=0.67\pm 0.10$). The
reason for the incorrect numerical result reported initially is the
small system sizes we studied, which did not allow us to reach the
asymptotic regime.

\vspace*{1cm} We thank M.E.J.~Newman and D.J.~Watts [3] and A.~Barrat
[4] for alerting us to the possibility of an error on our estimate of
$\tau$. We also thank M.~Argollo de Menezes for directing us to the
breadth-first search algorithm.

\vspace*{1cm}
%%%%%%%%%%%%%%%%%%%%%%%%%%%%%%%%%%%%%%%%%%%%%%%%%%%%%%%%%%%% References

[1] J. van Leeuwen, Ed., {\it Handbook of Theoretical Computer
Science. Volume A: Algorithms and Complexity}
(Elsevier, Amsterdam, 1990), p. 539.

[2] We found the LEDA libraries very useful and efficient
{\rm [http://www.mpi-sb.mpg.de/LEDA/leda.html]}.

[3] M.E.J.~Newman and D.J.~Watts, cond-mat/9903357.

[4] A.~Barrat, cond-mat/9903323; A.~Barrat and M.~Weigt,
cond-mat/9903411.

%%%%%%%%%%%%%%%%%%%%%%%%%%%%%%%%%%%%%%%%%%%%%%%%%%%%%%%%%%%

\begin{figure}
\narrowtext
\centerline{
\epsfysize=0.8\columnwidth{\epsfbox{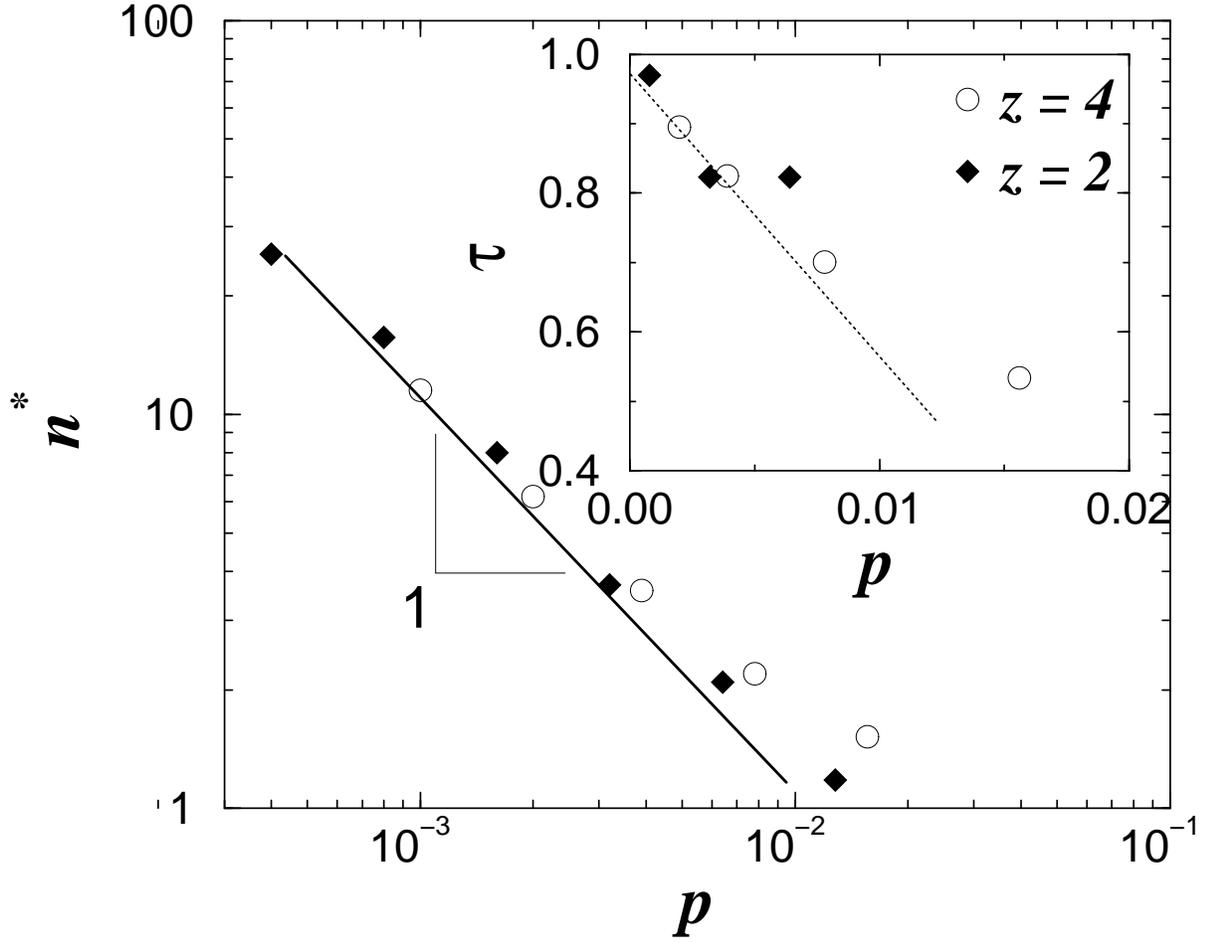}}
}
\vspace*{0.5cm}
\caption{Loglog plot of $n^*$ vs $p$ for system sizes up to $5500$ and
for $z=2$, 4. Note that the curvature of $n^*(p)$ in the loglog plot,
which gives us a local estimate of $\tau$, is increasing as $p$
decreases. In the inset, we show that $\tau$ approaches $1$ as $p \to
0$. Our new estimate of $\tau$ is $0.97\pm 0.05$, consistent with the
value 1 given by a simple scaling argument.}
\end{figure}

\end{document}